Doped and structured silica optical fibres for fibre laser sources

I. Kasik[1,*], I. Barton[1], M. Kamradek[1], O. Podrazky[1], J. Aubrecht[1], P. Varak[1], P. Peterka[1], P. Honzatko[1], D. Pysz[2], M. Franczyk[2], R. Buczynski[2]

[1] *Institute of Photonics and Electronics of the Czech Academy of Sciences, Chaberska 1014/57, Praha 8, 182 51, Czech Republic*

[2] *Lukasiewicz Research Network – Institute of Microelectronics and Photonics*
*Al. lotnikow 32/46, 02-668 Warsaw, Poland*

[*] *corresponding author kasik@ufe.cz*

Abstract
Specialty optical fibres, usually the silica-based ones doped with rare-earth ions, have been heart of fibre amplifiers and lasers spread thanks to work of team of Sir David N. Payne started in 1980-ies of 20th century. Wavelength of their emission depends on used rare earth, on glass matrix in which the rare earths are incorporated, on fibre structure in macro, micro and nano scale, and fibre laser architecture. Usually, fibre lasers are operated at single wavelength. A typical example is an erbium fibre laser (erbium ions in modified silica glass) operating around 1550 nm or ytterbium fibre laser (ytterbium ions in modified silica glass) operating around 1060 nm. When erbium and ytterbium ions together are randomly distributed in a silica glass matrix and pumped at absorption band of ytterbium, laser emission is typically obtained only at 1550 nm (emission of erbium) thanks to energy transfer from ytterbium to erbium ions, supported by modification of silica glass matrix with phosphorous pentoxide. However, when erbium and ytterbium ions are specifically structured in micro or nano scale in the fibre core it is possible to obtain dual-wavelength laser operation with controlled output parameters. Such dual-wavelength operation with controlled output at 1042 nm and simultaneously at 1550 nm was demonstrated with structured core $Er^{3+}$ and $Yb^{3+}$-doped fibre. The proposed approach makes fabrication of active fibres emitting with controlled characteristics at more wavelengths possible.

Introduction
The first fibre laser was presented in the period of 1960-ies earmarked by the invention of the laser by Eli Snitzer [1-2]. Snitzer employed Nd-doped optical fibre as an active lasing medium. For the following two decades, however, this invention stayed without any significant research interest. However, in mid-eighties, the idea of fibre lasers and optical amplification was revisited thanks to the extraordinary effort and inventions from the Optoelectronic Research Centre (ORC), University of Southampton, by the team that included Prof Sir David N. Payne, which soon became its leader. They firstly pioneered work in the field of optical fibres [3] and in optical amplification for telecommunications, which enabled Internet as we know it today. Researchers from the ORC developed optical amplifier based on $Er^{3+}$-doped fibre whose emission at wavelength of 1550 nm corresponds with the minimum loss of telecom fibres produced at that time [4-5]. Tradition of advancement in the fields of optical communications, optical fibres, fibre lasers (summarized, e.g. in [6-7]), and related technologies (e.g., [8]) has been linked with the research activities of David Payne's team. Large number of outstanding researchers and scientists world-wide have followed these trends. Even these days, rare-earth (RE) doped fibres, fibre lasers, and amplifiers keep representing key topics in leading scientific conferences and workshops. The authors of this paper would like to present their recently achieved results as a tribute to this fascinating inspiration.

Passive low-loss telecommunication fibres used for transmission have been produced by some of chemical vapor deposition methods like Modified Chemical Vapor Deposition (MCVD) [9], Plasma Chemical Vapor Deposition [10], Vapor Axial Deposition (VAD) or Outside Vapor Deposition (OVD) [11]. These methods are based on precursors in liquid state ($SiCl_4$, $GeCl_4$ etc.). However, starting materials for deposition of RE ions inevitable for fibre amplifiers and lasers are available mostly in solid state. Therefore, the methods of fibre fabrication had to be significantly modified. The first method developed at the ORC was based on controlled evaporation of chlorides of REs from a chamber placed at the inlet of substrate tube in the MCVD process [12-13]. Then a solution-doping method was developed and presented [14]. In this approach, porous core layer of silica soot particles is deposited at first by the MCVD method, then a solution containing salts of REs is applied, then solvent (water or alcohol) is evaporated, and finally doped core layer is sintered into glassy state. The first special fibres were doped with $Nd^{3+}$ [12], [15] and with $Er^{3+}$ ions [13]; germanium dioxide or phosphorus pentoxide were usually added into core matrix to achieve proper refractive index of the fibre core.

These results inspired number of scientists and launched global boom in research and development of RE-doped special optical fibres [11], [16]. A significant part of the research and development stayed focused on $Er^{3+}$-doped fibres, fibre amplifiers and fibre lasers summarized at [17-22]. The highest output power of 656 W achieved with $Er^{3+}$-doped fibres with a multimode operation has recently been reported [23-24].

Together with this research, fresh interest for variety of emission wavelength of laser sources raised and led to investigation of silica optical fibres doped with variety of RE (overview e.g., by Kirchhoff [25]), typically doped with $Yb^{3+}$, $Tm^{3+}$ and $Ho^{3+}$.

One of the first fibre lasers based on silica optical fibre doped with $Yb^{3+}$ and emitting around 1060 nm was demonstrated by Hanna [26]. Effort for increase of better efficiency (slope efficiency - SLE) and output power led to improving of ytterbium fibre laser to 17 mW of output power and 40 % slope efficiency [27]. Discovery of double-clad fibres led to significant increase of both parameters – 80% SLE was achieved by Pask [28], 90% SLE was achieved by Kurkov [29], then 1.36 kW output power and 83% SLE was achieved by Jeong [30]. Ytterbium-doped fibres were used in Q-switched fibre lasers [31] as well as in fibre-rod type fibre lasers and amplifiers of high-quality performance [32]. Recent progress of ytterbium fibre lasers was accompanied by fascinating increase of output power from kW-class to tens or hundreds of kW [7], [33].

Fibre lasers based on silica fibres doped with $Tm^{3+}$ with emission at "eye-safe" region around 1900 – 2000 nm was studied from 1990-ies [34]. Tunability of operating wavelength, increase of SLE (>50%) and output power were studied [35]. A comprehensive review on this topic did Jackson [36]. Progress of output power and SLE to multi-100 W scale and SLE >90% was achieved [37] and stopped at around this level (1 kW) [38-41]. Overheating which represents limitation of CW thulium-doped fibre lasers till these days has been studied by [42-43].

One of the first fibre lasers based on silica optical fibre doped with $Ho^{3+}$ emitting around 2100 nm was presented by Hanna [44]. Progress of SLE to 42% and 45.5% was achieved much later [45] and [46], respectively. Review on holmium-doped fibre lasers was performed e.g., by Hemming [47]. Fibre amplifier with peak gain of 25 dB at 2040 nm and with a 15 dB gain window spanning the wavelength range 2030 – 2100 nm was achieved [48]. Holmium-doped all-fibre laser pumped at 1125 nm and oscillating at around 2050 nm with total SLE 13% was demonstrated [49]. Tuneable holmium fibre laser with a maximum SLE of 58% at 2050 nm and 27% at 2200 nm with a total output power 8.9 W has recently been demonstrated [50]. The up-to-date status of holmium-doped fibre lasers includes cladding-pumped laser with 400 W of output power (40% SLE) [51] or core-pumped sources with output in a range of tens of Watts obtained with SLE above 80% [52-53].

Lack of suitable pumping sources in the past or demand for enhancement of laser properties (SLE, output power) led also to investigation of fibres codoped with more REs exploiting potential effect of energy transfer between RE ions. A typical example are fibres usually denoted as Er/Yb which are doped with $Er^{3+}$ ions emitting at around 1550 nm and sensitized with $Yb^{3+}$ ions. First Er/Yb fibre lasers were usually pumped at 1064 nm by YAG lasers available at that time; output power of 4 mW and 7 mW were achieved at the beginning [54]. Threshold of 5 mW and SLE 8.5% was achieved by [55], thresholds of 13.5 mW and 8.5 mW, SLE 3% and 5% and power 0.75 and 0.33 mW were achieved by [56]. Er/Yb fibres and fibre lasers represented mainstream of research in this field in 1990-ies. A comprehensive study of fibre glass material and theoretical modelling of Er/Yb fibre lasers was studied and presented [57]. Q-switched fibre lasers were demonstrated – with 70 ps short pulses [58], with 7ps short pulses of 200 W output power [59], with 2 ps short pulses of 10 mW output power [60]. Er/Yb fibre amplifier with +24.6 dBm signal gain was demonstrated [61] and later with +34.9 dB signal gain [62]. Fibre laser of 19mW output power and of SLE 55% was demonstrated [63] and later generating of >1 W output power was achieved [64]. Recent development of 345 W output power of Er/Yb all-fibre laser was reported [65], [24].

Dual-wavelength operation of Er/Yb double-clad non-structured fibre (i.e., without of controlled output power) in specific task of difference frequency generation was reported by Krzempek [66]. Partial study and results on dual-wavelength fibre lasers based on active fibres with structured cores were presented at conferences CLEO/Europe-EQEC [67], SPIE – Optics and Optoelectronics [68] and Photonics West 2024 [69]. Studies of doping with $Tm^{3+}/Ho^{3+}$ for optical generation were performed as well – in silicate fibres [70] or in silica fibres [71-72].

Number of methods and techniques have been elaborated to be able to prepare such materials and fibres. Among others: vapor phase chelate delivery method [73-76], flash-condensation technique [77], aerosol-based method [78], halide-evaporation method [79], molten-core method [80], powder-based methods [81-83] and some others. Research team of authors of this paper were also inspired by this stream [84-85].

Interest for gradual enhancement of higher output power of fibre lasers and high-power fibre lasers led to development of novel laser arrangements, modification of fibre structures and increase of concentration of REs in fibre cores. Unfortunately, REs is not miscible with silica glass and cause clustering and phase separation even at low concentration (above around 200 ppm of RE) [86]. Therefore, it was necessary to find suitable modificators of glass matrix which would be transparent in near-infrared spectral region, would dissolve REs, and would be miscible with silica glass. Binary core matrices (like $Al_2O_3$-$SiO_2$ [87-90], $P_2O_5$-$SiO_2$ [91-92]) and ternary core matrices (like $GeO_2$-$P_2O_5$-$SiO_2$ [93-94], $Al_2O_3$-$P_2O_5$-$SiO_2$ [95-100]) and more-component compositions have been investigated for years with the aim to increase final content of modifying oxides in silica glass to diminish phase separation and so to increase of RE content in core matrix. Glass ceramics materials have been prepared alternatively as well [101].

Current trends in this field are adherent to the introduction of nanotechnologies. Modification of core matrices by metallic or semiconductor nanoparticles for enhancement of laser performance was tested at first [102-103]. Later, implementation of ceramics nanoparticles was investigated [104-105]. Novel relevant methods were developed like direct particle deposition [106], nanophase separation [107-110] nanoparticle technique extending the MCVD [111]; several overviews have been summarized [112-114]. Authors of this paper supported inauguration of this trend by elaboration of nanoparticle-doping method and presenting it first at [115], later [116-117]. This method has led to production of doped fibres of parameters appreciated by established research groups like [118-119].

A breakthrough stack-and-draw method [120] introduced a powerful tool for nanostructuring of optical fibres at the beginning of Millenium. This concept was originally developed for making of endoscopes [121] and its later implementation in the field of RE-doped optical fibers made fabrication of large mode area (LMA) structures possible. This concept investigated theoretically and experimentally by other research groups potentially leads to fibres of almost arbitrary design, gain and refractive index profile design [122]. So, it makes fabrication of active fibres emitting controllably at more wavelengths possible.

In this paper we review our results in preparation, characterization, and performance of $Er^{3+}$ and $Yb^{3+}$-doped (nano)structured core optical fibres operated as gain medium for dual-wavelength fibre laser emitting at around 1000 nm thanks to ytterbium-doped regions and at around 1500 nm thanks to erbium-doped regions.

$Er^{3+}$ and $Yb^{3+}$-doped silica fibres with structured core

Design, fabrication and characterization of fabricated silica-based $Er^{3+}$ and $Yb^{3+}$ -doped fibres emerged from experience acquired with phosphate-based fibres for dual-wavelength operation [123]. These fibres were fabricated by doubled (repeated two times) stack-and-draw process which reduced size of doped regions to nano- scale and thousands of such "nanorods" (c.a. 160 nm in diameter each) formed the single-mode fibre core with effective step-index refractive index profile. Following this concept, silica-based fibres presented in this paper were fabricated by doubled stack-and-draw process leading to "nanostructured core fibres" and by single stack-and-draw process leading to "structured core fibres".

Experimental

Initial preforms of core composition $Er^{3+}$-$Al_2O_3$-$SiO_2$ and $Yb^{3+}$-$Al_2O_3$-$SiO_2$ were prepared by nanoparticle-doping method [115] which is a specific extension of the MCVD process [9]. $ErCl_3$ and $YbCl_3$ (99.998%, Aldrich) and $Al_2O_3$ nanoparticles (<50 nm, Sigma-Aldrich No.544833-506) were used for the experiments. Initial modelling of suitable Er/Yb ratio of final fibre core and active length was performed [68] taking into account characteristics such as refractive-index profile and chemical composition of prepared initial preforms.

Fabricated MCVD preforms were uniformly etched by hydrofluoric acid to achieve suitable core-silica ratio predicted by the initial numerical model. Then the MCVD preforms were elongated at drawing tower to rods of proper diameter of 390 μm.

Nanostructured core fibre (Fibre#1) was drawn by doubled stack-and-draw process. It means that the first preform was assembled from 91 rods of elongated initial MCVD preforms and then the obtained 19pcs of elongated stack were again used for assembling the final preform. In this way, 1729 RE-doped nano-spots of $Er^{3+}$/$Yb^{3+}$ ratio ~40/60 was achieved in final fibre of 6 μm diameter single-mode core. Structured core fibre (Fibre#2) was drawn by single stack-and-draw process. In this case, 7 rods of elongated initial MCVD preforms with diameter of 390 μm were arranged into hexagonal stack of five $Er^{3+}$-doped rods and two $Yb^{3+}$-doped rods ($Er^{3+}$/$Yb^{3+}$ ratio of ~30/70). The stack was loaded into sleeving silica tube (F300, Heraeus) and such preform was drawn at a temperature of 1940°C into a fibre of diameter 125μm, core diameter 6 μm and coated with conventional UV-curable acrylate coating DeSolite 3471-3-14. The technology concept is described at Fig. 1.

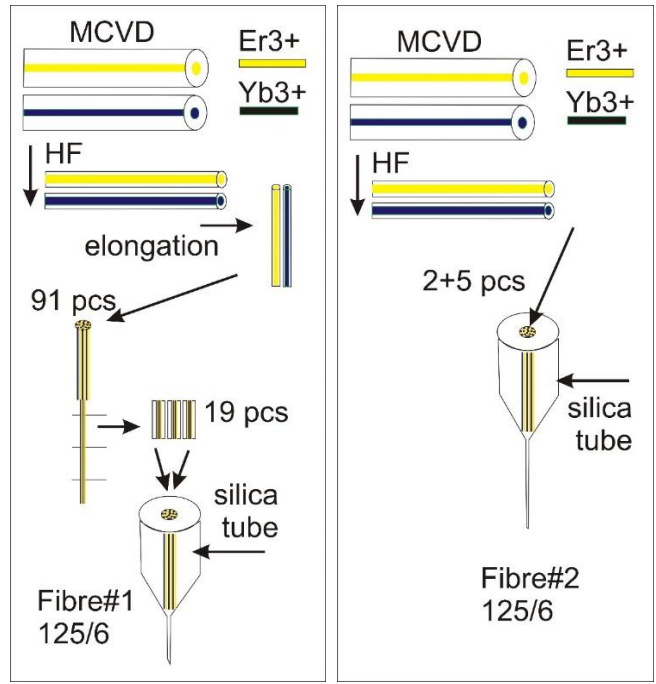

Fig.1 Scheme of fabrication of a)-left nanostructured $Er^{3+}$ and $Yb^{3+}$-doped silica fibre (Fibre#1) and b)-right structured core $Er^{3+}$ and $Yb^{3+}$-doped silica fibre (Fibre#2).

Nanostructured-core $Er^{3+}$ and $Yb^{3+}$ -doped phosphate optical fibre compared finally to performance of prepared silica fibres was fabricated separately by double stack-and-draw process from rods from in-house active phosphate glass doped with $Yb_2O_3$ or with $Er_2O_3/Yb_2O_3$ [123].

Initial preforms prepared by the MCVD process extended with nanoparticle doping were characterized by refractive index profile (RIP) and by local chemical composition by Electron Microprobe Analysis (EMA). Optical profiler (Photon Kinetics A2600) and EMA profiler (JEOL JXA-8230) were used. Fabricated fibre was characterized by refractive-index profile (IFA-100, Interfibre Analysis Inc.), by spectral attenuation by cut-back method, by optical microscopy (Olympus BX51) or Scanning Electron Microscope (Lyra 3GM, Tescan) (SEM) and by lifetime measurements The fluorescence decay curves were measured using setup composed of an Agilent 3512B pulse generator and ILX lightwave power source, EM4 P161-600-976 diode emitting at 976 nm as an excitation source, Thorlabs FELH1000 or FELH1150 optical isolator, PDA36A Thorlabs photodetector or S2386-18K Hamamatsu Si photodiode detector and Teledyne Lecroy HDO6034 oscilloscope. The measured fibres were approximately 1 mm long and the emission was detected from the side, to suppress the influence of amplified spontaneous emission (ASE). Details of lifetime measurements are described [124-126]. Lasing characteristics of the fabricated fibres were determined in Fabry-Perot configuration with a pumping source operating at 974 nm (Lumics) with a maximum output power of 450 mW. The laser cavities for both erbium and ytterbium laser were formed by two single high-resolution FBG (HRFBGs reflecting at 1064nm and 1561nm) or by one single HRFBG (reflecting at 1042nm and 1550nm). The active fibre which was perpendicularly cleaved at the output end to get a low-reflectivity mirror through Fresnel reflection. The optical filters (Thorlabs, FELH1000 or FELH1150) with absorption edges at 1000 nm and 1150 nm were gradually placed in a forward direction before the thermopile power detector (Gentec, XLP12-3S-H2-D0) to separate the pump and individual signal beams. The scheme of fibre laser setup used for characterization is depicted at Fig.2.

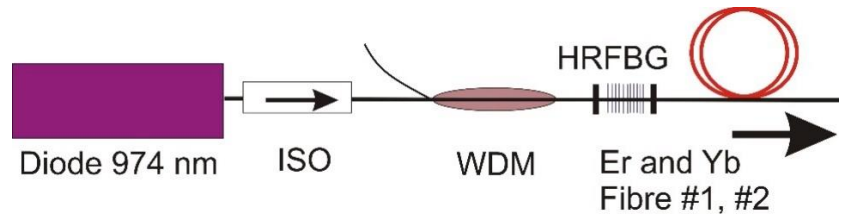

Fig. 2 Fibre laser setup used for characterization.

Results and discussion

Preforms of diameter of around 9.5 mm and of length of around 350 mm were prepared without visible inhomogeneities, phase separation, bubbles, or clusters.

Assembling of preforms for stack-and-draw fabrication method requires high number of incoming $Yb^{3+}$ and $Er^{3+}$-doped rods and MCVD preforms. Therefore, a set of more than 30 initial MCVD preforms was prepared and characterized. Attention was focused on longitudinal homogeneity of individual preforms and on repeatability of preform properties. Initial MCVD preforms were prepared with variation of RIPs within ~10%. Therefore, RIPs of two typical preforms ($Er^{3+}$-doped and $Yb^{3+}$-doped) were chosen to show at Fig. 3. No central dip can be seen. Such character of RIP corresponds to doping of core matrix with non-volatile $Al_2O_3$. A smooth character of RIPs corresponds to high quality (transparency) and radial homogeneity of preform cores without phase separation or imperfections on core-silica substrate boundary. A minimum difference between RIPs measured at the middle of the preforms and at their end can be observed as evidence of high longitudinal homogeneity of each preform. A small difference between RIPs of individual preforms corresponds to satisfactory level of repeatability of the MCVD fabrication process. High level of doping of initial preforms can be expected from relatively high maximum refractive index difference of cores (0.023 and 0.025); core diameter of preforms (FWHM) of around 1.34 mm can be seen.

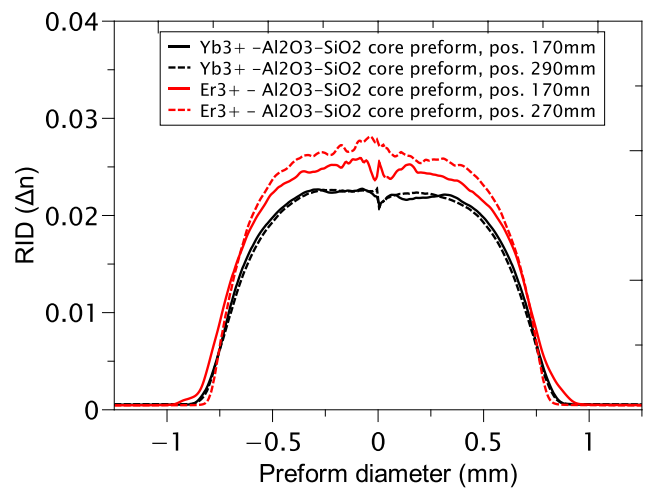

Fig. 3 RIPs of typical initial MCVD preforms doped with $Yb^{3+}$ (black) or $Er^{3+}$ (red) and $Al_2O_3$ used for structuring, measured at the end and middle of the preforms.

A maximum content of around 10 mol% of $Al_2O_3$, of 3500 mol ppm $Er^{3+}$ and of 5000 mol ppm $Yb^{3+}$ was determined by EMA analysis (Fig. 4) in initial MCVD preforms. Desired high Al/RE ratio [117] (60 and 40, respectively) can also be seen from this chemical analysis. EMA analyses of stacks were not performed since individual rods are assembled without longitudinal fixing and making of sample for analysis is destructive in such case. Lateral resolution of EMA does not allow local analysis of core composition of prepared fibres; SIMS analyses requested so far did not lead to satisfactory results as well.

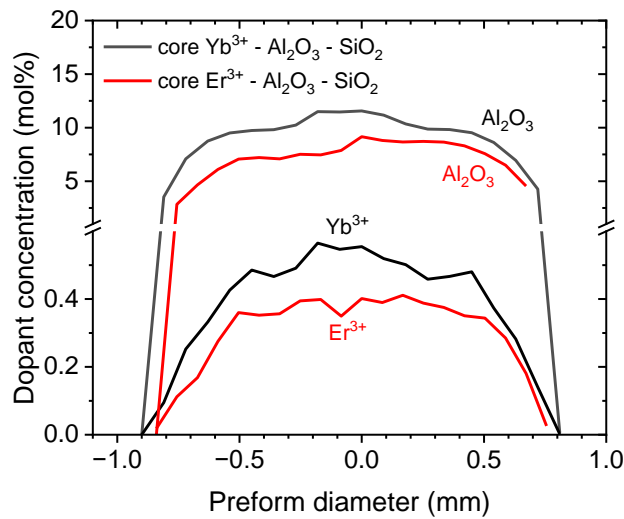

Fig. 4 Distribution of $Al_2O_3$, $Yb^{3+}$ and $Er^{3+}$ dopants in typical initial MCVD preforms determined by EMA.

RIPs of nanostructured and structured core fibres drawn from preforms assembled from 1729 rods (Fibre#1) and from 7 rods (Fibre#2) can be seen in Fig. 5. Contrasting character of RIP of Fibre#2 in comparison to RIP of conventional initial MCVD preforms and to RIP of Fibre#1 can be observed. RIP of Fibre#2 is not smooth because the fibre core region is composed from stack of 2+5 rods elongated from initial MCVD preforms, each containing its own doped core (Fig. 1 - right). RIP of nanostructured Fibre#1 is smooth because structuring of the core is in nanoscale (Fig. 1 - left) and so the discrete character of the refractive index distribution cannot be observed in visible spectral range. Overall, the effective refractive index is much lower than that of initial MCVD preforms doped cores; it corresponds to averaging of refractive index of initial cores and silica glass claddings. Considering average refractive index of Fibre#2 (~1.463) and its diameter (FWHM of around 6 µm), cut-off wavelength can be estimated of around 1050 nm proving single-mode character of this fibre. The size of the fibre core (as well as of the core of Fibre#1 below) was not determined of high accuracy since the core was inherently not angularly homogeneous and method of such measurements (modelling) would deserve a separate presentation. Nevertheless, presented approach can be considered sufficient for estimating of cutoff and single-mode regime of performance of built-up fibre laser.

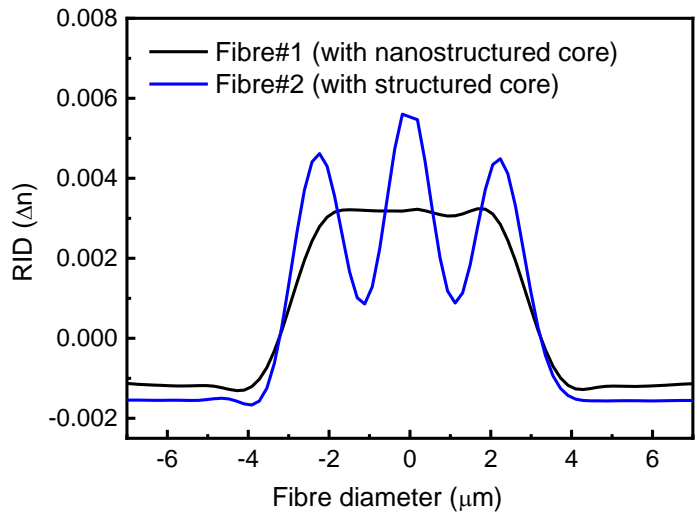

Fig. 5 RIP of $Er^{3+}$ and $Yb^{3+}$-doped fibres with nanostructured core (Fibre#1) and structured core (Fibre#2) measured with IFA-100.

The cross sections of $Er^{3+}$ and $Yb^{3+}$-doped fibres with nanostructured core (Fibre#1) and structured core (Fibre#2) can be seen in Fig. 6. The cross section of Fibre#1 characterized by SEM can be seen in Fig. 6a. Mapping by SEM of high spatial resolution was chosen for this characterization to provide better chance to observe imperfections like potential inhomogeneities or cluster formation inside the nanostructured core. However, even this resolution does not allow us to observe structure of core material in nanoscale details. So, distinct nanostructure of the nanostructured core was not observed, only shape of the core of low contrast to the rest of the fibre is visible at the middle of the figure. Black circular spot on the left side of the figure is an artefact, an imperfection caused by breaking of brittle fibre during sample preparation can be observed on the right side of the figure. Total fibre diameter is of 125 μm; diameter of core area of Fibre#1 (of around 6 μm) corresponds to dimensions observed from RIP of fibre. No structuring of the fibre core can be observed at used scale rendering good comparison of structured and nanostructured fibre.

The cross section of the Fibre#2 was characterized by optical microscopy (transmission arrangement) as can be seen in Fig. 6b. Mapping by optical microscopy of spatial resolution in microns can sufficiently display the structured core since size of individual spots (originating in assembled rods) is large enough (above Abbe´s limit) in this case. Fibre#2 is circular and radially symmetric; deviation from circularity (like D-shape) is an imperfection caused by breaking of brittle fibre during sample preparation. Total fibre diameter is of 125 μm; diameter of area of Fibre#2 corresponds to dimensions observed from RIP of fibre. Seven bright circles forming the structured core fibre core can be observed. They correspond to doped cores of elongated initial MCVD preforms and they are light guiding (bright) thanks to higher refractive index than their surroundings. Darker areas correspond to silica glass claddings in the vicinity of these cores and to silica glass oversleeving tube surrounding the structured fibre core. No holes or cavities are present.

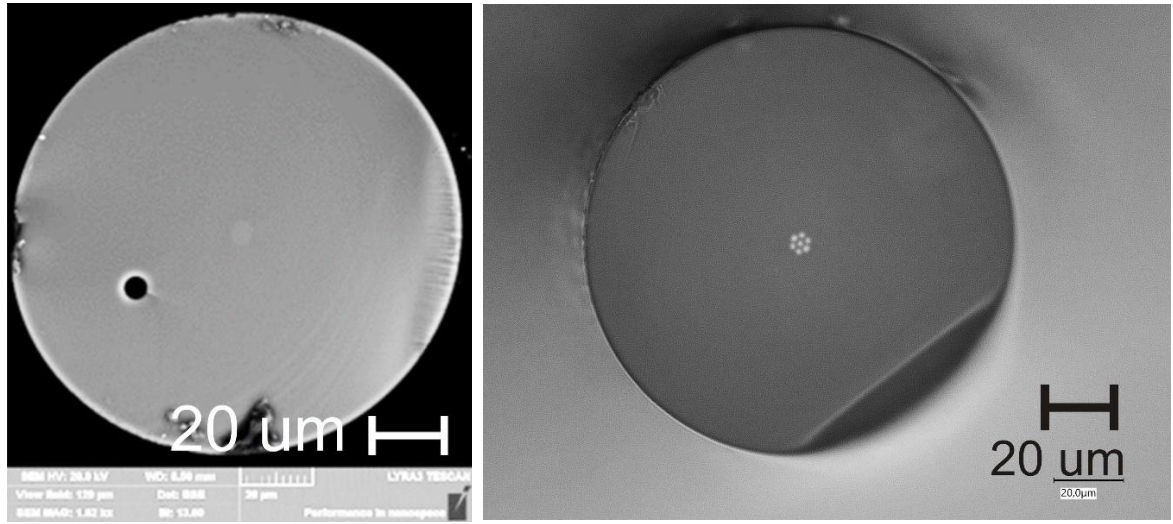

Fig. 6 Cross section of $Er^{3+}$ and $Yb^{3+}$ doped fibres a) SEM of Fibre#1, core composed from 1729 initial rods, b) Micro photo of Fibre#2, core composed from 7 initial rods.

Optical losses of Fibre#1 can be seen at Fig. 7. Absorption bands of $Yb^{3+}$ and $Er^{3+}$ are depicted in Fig 7a, background losses in near infrared region are depicted in Fig. 7b. Optical losses of maximum of absorption band of $Yb^{3+}$ at 978 nm were determined of ~63 dB/m, maximum of absorption band of $Er^{3+}$ at 1531 nm was determined of ~14 dB/m. Minimum background losses were observed at 1278 nm and at 1310 nm of around ~0.23 dB/m.

These characteristics are close to the values determined with fibre with structured core. Optical losses of Fibre#2 were determined: maximum of absorption band of $Yb^{3+}$ at 978 nm ~70.14 dB/m, maximum of absorption band of $Er^{3+}$ at 1531 nm ~12.4 dB/m, minimum background losses at 1220 nm and at 1310 nm (~0.15 dB/m). Both values of minimum background losses (Fibre#1 as well as Fibre#2) were determined as minimum background losses observed (not determined at a particular wavelength). It is difficult to interpret slightly lower background losses observed in structured fibre. Probably it can be attributed to operations during assembling of preforms and level of their purity (like in case of fabrication of microstructured fibres). Single stack-and-draw process gives lower opportunity for introducing of impurities like dust into preform than double stack-and-draw process, so final losses can be lower.

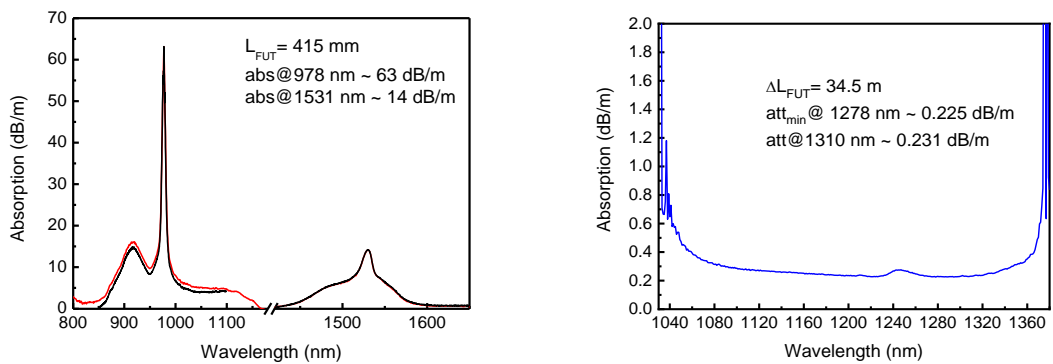

Fig. 7 Optical losses of Fibre#1 a) absorption bands of $Er^{3+}$ and $Yb^{3+}$, b) background losses.

Fluorescence lifetimes of fibres were determined. Fig. 8 describes fluorescence lifetimes of $Er^3$ (a) and $Yb^{3+}$ (b) ions of Fibre#1. The fluorescence lifetime of the $Er^{3+}$ ion in the 4I13/2→4I15/2 transition was 10.13 ms, the $Yb^{3+}$ ions in the 2F5/2→2F7/2 transition exhibited a lifetime of 0.763 ms.

The fluorescence lifetime of the $Er^{3+}$ ion in the 4I13/2→4I15/2 transition of Fibre#2 was 10.1 ms, which is in good agreement with values typically found in $Er^{3+}$-doped silica fibres [126]. The $Yb^{3+}$ ions in the 2F5/2→2F7/2 transition in Fibre#2 exhibited a lifetime of 0.84 ms.

From the comparison of fluorescence lifetimes of Fibre#2 fabricated by single stack-and-draw process and Fibre#1 fabricated by doubled stack-and-draw process can be seen that lifetime of $Er^{3+}$ ions stay identical for both fibres (of around 10.1 ms) while lifetime of fluorescence of $Yb^{3+}$ ions slightly differ. This discrepancy led us to formulation of hypothesis that lifetimes of individual rare earths depend on history of their thermal processing in glass matrix. Details of optical characterization of the fibres and results related to the hypothesis are described in [124, 126].

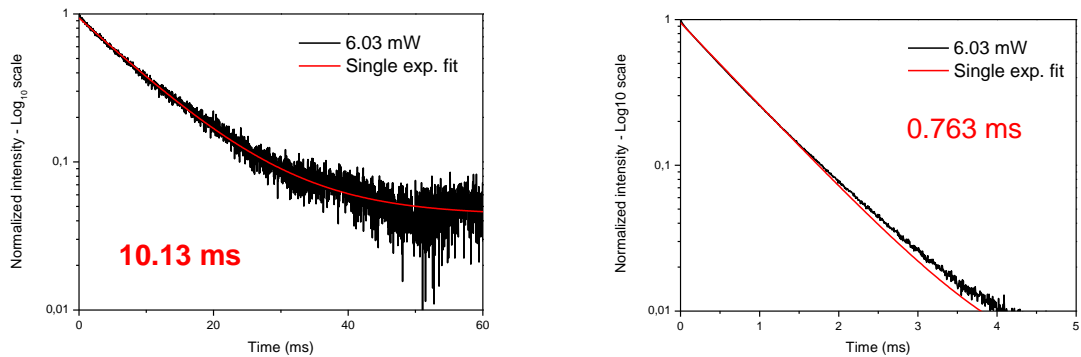

Fig. 8 Fluorescence lifetimes of $Er^{3+}$ ions (a) and $Yb^{3+}$ ions, (b) embedded in Fibre#1.

Finally, laser performance of the fabricated fibres was characterized (Fig. 9).

Active length of structured as well as nanostructured fibre was one of important characteristics of the fibre laser (like Er/Yb ratio, input power etc.). Active fibre length used in the experiments was optimized in agreement with initial results of fibre-laser modelling [68]. Since part of the fibre core is not doped with rare earth ions (thanks to rest of silica cladding of stacked rods), either active fibre length shall be longer in comparison to fibre lasers based on fibres prepared from preforms fabricated by conventional solution-doping method, or, the rare-earth concentration would be increased, in the case it is technologically feasible. The part of work dealing with fibre laser modelling and characterization is quite extensive and it will be published separately; here the part proving the dual-wavelength fibre laser performance is presented.

Emission of only $Er^{3+}$ ions at around 1550 nm (Fig. 9a) was observed from Fibre#1 (with nanostructured core); SLE of this lasing was of around 21%. No emission of $Yb^{3+}$ ions was observed (in a range of 0-220 mW of launched pump) even when input power was significantly increased or testing laser setup (HRFBG) modified. It can be explained by unwanted diffusion of dopants in the core which is much higher when the fibre was prepared by repeated double stack-and-draw thermal process (nanostructured) in comparison to the core fibre structured by only one stack-and-draw thermal process (Fig. 1). This effect is still under investigation. In the case of Fibre#2 (with structured core), two distinct lasing peaks at the emission spectrum measured at forward direction of comparable (controlled) output power can be seen (Fig. 9b). Emission at shorter wavelength of 1042 nm can be attributed to $Yb^{3+}$ ions (SLE 30%), emission at longer wavelength of 1550 nm can be attributed to $Er^{3+}$ ions (SLE 19%) [69]. This can be considered as a proof of dual-wavelength performance of the fibre laser based on $Yb^{3+}$ and $Er^{3+}$-doped structured core silica fibre.

Shift of $Yb^{3+}$ emission of Fibre#2 to shorter wavelength was caused by different HRFBG used for lasing characteristics measurements (cavities were formed by two single HRFBGs, reflecting at 1064nm and 1561nm, or by one single HRFBG, reflecting at 1042nm and 1550nm, simultaneously).

The results obtained are in agreement with dual-wavelength operation of $Yb^{3+}$ and $Er^{3+}$-doped phosphate nanostructured core fibre (Fig. 9c), in which the separation of $Yb^{3+}$ and $Er^{3+}$ doped regions was expected, as stated in [123]. Different character of emission and length of fibre laser based on phosphate-glass fibre in comparison to silica-based fibre lasers can be explained by different character of glassy matrix which influences absorption and fluorescence properties of active rare-earth ions.

This behaviour contrasts, thanks just to fibre core structuring, with performance of conventional fibre lasers based on Er/Yb fibres prepared by conventional solution-doping method and emitting at single wavelength of around 1550 nm thanks to energy transfer from $Yb^{3+}$ to $Er^{3+}$ ions [57], [127].

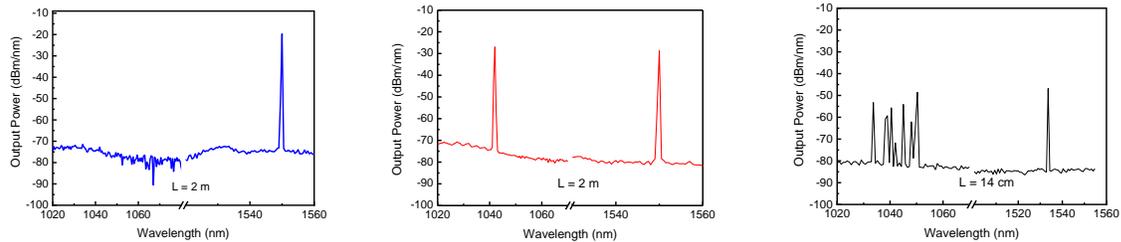

Fig. 9 Laser emission of $Er^{3+}$ and $Yb^{3+}$-doped fibres a) Fibre#1 (with nanostructured core), b) Fibre#2 (with structured core), and c) nanostructured phosphate fibre.

Conclusions

$Er^{3+}$ and $Yb^{3+}$-doped nanostructured and structured core optical fibres were fabricated, characterized, and examined for fibre laser operation. Dual-wavelength operation with controlled output at 1042 nm and simultaneously at 1550 nm was observed with structured core $Er^{3+}$ and $Yb^{3+}$-doped fibre. This behaviour contrasts just to fibre core structuring with performance of conventional fibre lasers based on Er/Yb optical fibres. The proposed approach in general makes fabrication of active fibres emitting controllably at more wavelengths possible.

Details of modelling, characterization and performance of such fibre and fibre laser exceeds scope of this paper and will be published separately. Acquired data will serve as inputs for next progress of numerical modelling of such fibre lasers. Better understanding of fluorescence lifetime changes and diffusion (spatial separation of doped regions) during high temperature processes of fibre fabrication will be in focus of next research. An impact to research and development of fibre lasers emitting at more wavelengths with controlled characteristic can be expected.

Acknowledgement
This work was supported by the Grant Agency of CR (contract 21-45431L), Narodowe Centrum Nauki (OPUS LAP 2020/33/IST7/02143) and co-funded by the EU project under the project LasApp CZ.02.01.01./00./22_008/0004573.

CRediT authorship contribution statement

I. Kasik: Conceptualization, Funding acquisition, Investigation, Project administration, Supervision, Writing – original draft, Writing – review &editing. I. Barton: Investigation, Writing – review &editing. M. Kamradek: Investigation, Writing – review &editing. O. Podrazky: Investigation. J. Aubrecht: Investigation, Writing – review &editing. P. Varak: Investigation, Writing – review &editing. P. Peterka: Conceptualization, Investigation, Supervision. P. Honzatko: Writing – review &editing. D. Pysz: Investigation. M. Franczyk: Investigation, Writing – review &editing. R. Buczynski: Conceptualization, funding acquisition, Project administration, Supervision.

Data availability statement

The data supporting the results of this study are available in [128].

References

[1] E. Snitzer, Optical Maser Action of $Nd^{3+}$ in a Barium Crown Glass, Phys. Rev. Lett. 7 (1961) 444. https://doi.org/10.1103/PhysRevLett.7.444

[2] C.J. Koester, E. Snitzer, Amplification in a fibre laser, Appl. Opt. 3 (1964) 1182-1186. https://doi.org/10.1364/AO.3.001182

[3] D.N. Payne, W.A. Gambling, New silica-based low-loss optical fibre, Electron. Lett. 10 (1974) 289 – 290. https://doi.org/10.1049/el:19740231

[4] R.J. Mears, L. Reekie, I.M. Jauncey, D.N. Payne, Low-noise erbium-doped fibre amplifier operating at 4.54 μm, Electron. Lett. 23 (1987) 1026-1028. https://doi.org/10.1049/EL%3A19870719

[5] S.G. Grubb, W.F. Humer, R.S. Cannon, T.H. Windhorn, S.W. Vendetta, K. L. Sweeny, P.A. Leilabady, W.L. Barnes, K.P. Jedrzejewski, J.E. Townsend, +21 dBm Erbium power amplifier pumped by a diode-pumped Nd:YAG laser, IEEE Photon. Technol. Lett. 4 (1992) 553-555. https://doi.org/10.1109/68.141965

[6] D.J. Richardson, J. Nilsson, W.A. Clarkson, High power fibre lasers: current status and future perspectives (invited), J. Opt. Soc. Am. B 27 (2010) B63-B92. https://doi.org/10.1364/JOSAB.27.000B63

[7] M.N. Zervas, C. Codemard, High power fibre lasers: a review (invited), 2014. IEEE J. Selected topics in quantum el. 20, 0904123. https://doi.org/10.1109/JSTQE.2014.2321279

[8] A.S. Webb, A.S. Boyland, R.J. Standish, S. Yoo, J.K. Sahu, D.N. Payne, MCVD in-situ doping process for the fabrication of complex design large core rare-earth doped fibres, J. Non-Cryst. Solids 356 (2010) 848-851. https://doi.org/10.1016/j.jnoncrysol.2010.01.008

[9] S.R. Nagel, J.B. MacChesney, K.L. Walker, An overview of the Modified Chemical Vapor Deposition (MCVD) process and performance, IEEE J. Quantum. El. Q-17 (1982) 459-476. https://doi.org/10.1109/TMTT.1982.1131071

[10] P. Geittner, D. Kuppers, H. Lydtin, Low-loss optical fibres prepared by plasma-activated chemical vapor-deposition (CVD), Appl. Phys. Lett. 28 (1976) 645-646. https://doi.org/10.1063/1.88608

[11] A. Mendez, F.T. Moorse, Specialty Optical Fibres Handbook. 1st ed., Elsevier, 2007. ISBN 978-0-12-369406-5.

[12] S.B. Poole, D.N. Payne, M.E. Fermann, Fabrication of low-loss optical fibres containing rare-earth ions, Electron. Lett. 21 (1985) 737-738. https://doi.org/10.1049/el:19850520

[13] S.B. Poole, D.N. Payne, R.J. Mears, M.E. Fermann, R.I. Laming, Fabrication and characterization of low-loss optical fibres containing rare-earth ions, J. Lightwave Technol. LT-4 (1986) 870-876. http://doi.org/10.1109/JLT.1986.1074811

[14] J.E. Townsend, S.B. Poole, D.N. Payne, Solution-doping technique for fabrication of RE-doped optical fibres, Electron. Lett. 23 (1987) 329-331. https://doi.org/10.1049/EL%3A19870244

[15] J. Stone, C.A. Burrus, Neodymium-doped silica lasers in end-pumped fibre geometry, Appl. Phys. Lett. 23 (1973) 388-389. https://doi.org/10.1063/1.1654929

[16] R. Paschotta, Optical fibre technology, SPIE, 2010, ISBN 078-0-8194-8090-3.

[17] E. Desurvire, J.R. Simpson, Amplification of spontaneous emission in Er-doped single-mode fibres, J. Lightwave Technol. 7 (1989) 835-845. https://doi.org/10.1109/50.19124

[18] E. Desurvire, Erbium-doped fibre amplifiers: principles and applications, John Wiley & Sons, 1994. ISBN 0-471-58997-2.

[19] W.J. Minscalco, Erbium-doped glasses for fibre amplifiers at 1500 nm (invited), J. Lightwave Technol. 9 (1991) 234-250. http://dx.doi.org/10.1109/50.65882

[20] M.J.F. Digonnet, Rare-earth-doped fibre lasers and amplifiers, 2nd ed. Marcel Dekker, 2001. ISBN 978-0-8247-0458-2.

[21] K.S. Abedin, T.F. Taunay, M. Fishteyn, D.J. Digiovanni, V.R. Supradeepa, J.M. Fini, M.F. Yan, B. Zhu, E.M. Monberg, F.V. Dimarcello, Cladding-pumped erbium-doped multicore fibre amplifier, Opt. Ex. 20 (2012) 20191-20200. https://doi.org/10.1364/OE.20.020191

[22] P. Peterka, J. Vojtech, Optical amplification, in Handbook of radio and optical networks convergence, Springer, 2023. https://doi.org/10.107/978-981-33-4999-5_20-1

[23] H.Q. Lin, Y.J. Feng, Y.T. Feng, P. Barua, J.K. Sahu, J. Nilsson, 656 W Er-doped, Yb-free large-core fibre laser, Opt. Lett. 43 (2018) 3080-3083. https://doi.org/10.1364/OL.43.003080

[24] G. Nemova, 2024. Brief review of recent developments in fibre lasers, Appl. Sci. 14, 2323. https://doi.org/10.3390/app14062323

[25] J. Kirchhof, S. Unger, A. Swuchow, Fibre lasers: materials, structures and technologies (invited), Proc. Optical fibres and sensors for medical applications III, Vol 4957 (2003) 1-15. http://dx.doi.org/10.1117/12.498062

[26] D.C. Hanna, R.M. Percival, I.R. Perry, R.G. Smart, P.J. Suni, J.E. Townsend, A.C. Tropper, Continuous-wave oscillation of a monomode ytterbium-doped fibre laser, Electron. Lett. 24 (1988) 1111-1113. https://doi.org/10.1049/el:19880755

[27] J.R. Armitage, R. Wyatt, B.J. Ainslie, S.P. Craig-Ryan, Highly efficient 980 nm operation of an $Yb^{3+}$-doped silica fibre laser, Electron. Lett. 25 (1989) 298-299. https://doi.org/10.1049/el:19890208

[28] H.M. Pask, J.L. Archambault, D.C. Hanna, L. Reekie, P.St.J. Russel, J.E. Townsend, A.C. Tropper, Operation of cladding-pumped $Yb^{3+}$-doped silica fibre lasers in 1 μm region, Electron. Lett. 30 (1994) 863-865. https://doi.org/10.1049/el:19940594

[29] A.S. Kurkov, V.I. Karpov, A.Y. Laptev, O.I. Medvedkov, E.M. Dianov, A.N.Guryanov, S.A Vasilev, V.M. Paramonov, V.N. Protopopov, A.A. Umnikov, N.I. Vechkanov, V.G. Artyushenko, Y. Fram, Highly efficient cladding-pumped fibre laser based on an ytterbium-doped optical fibre and a fibre Bragg grating, Quantum Electronics 29 (1999) 516-517. http://dx.doi.org/10.1070/QE1999V029N06ABEH001520

[30] Y. Jeong, J.K. Sahu, D.N. Payne, J. Nilsson, Ytterbium-doped large-core fibre laser with 1.36 kW continuous-wave output power, Opt. Ex. 12 (2004) 6088-6092. https://doi.org/10.1364/OPEX.12.006088

[31] A.V. Kir'yanov, Y.O. Barmenkov, Self-Q-switched ytterbium-doped all-fibre laser, Laser Phys. Lett. 3 (2006) 498-502. https://doi.org/10.1002/lapl.200610039

[32] C.R. Smith, N. Simakov, A. Hemming, W.A. Clarkson, Thermally guided Yb-doped fibre-rod amplifier and laser, Applied Physics B 125 (2019) 32. https://doi.org/10.1007/s00340-018-7126-3

[33] www.ipgphotonics.com, 2024 (accessed 4 September 2024)

[34] T. Yamamoto, Y. Miyajima, T. Komukai, 1.9 μm Tm-doped silica fibre laser pumped at 1.57 μm, Electron. Lett 30 (1994) 220-221. https://doi.org/10.1049/el:19940135

[35] T.S. McComb, R.A Sims, C.C. Willis, P. Kadwani, V. Sudesh, L. Shah, M. Richardson, High-power widely tuneable thulium fibre lasers, Appl. Opt. 49 (2010) 6236-6242. http://dx.doi.org/10.1364/AO.49.006236

[36] S.D. Jackson, Towards high-power mid-infrared emission from a fibre laser, Nature Phot. 6 (2012) 423-431. https://doi.org/10.1038/nphoton.2012.149

[37] A. Sincore, J.D. Bradford, J. Cook, L. Shah, M.C. Richardson, High average power thulium-doped silica fibre lasers: Review of systems and concepts, IEEE J. Selected topics quantum el. 24 (2018) 0901808. http://dx.doi.org/10.1109/JSTQE.2017.2775964

[38] C. Gaida, M. Gebhardt, T. Heuermann, F. Stutzki, C. Jauregui, J. Limpert, Ultrafast thulium fibre laser system emitting more than 1 kW of average power, Opt. Lett. 43 (2018) 5853-5856. https://doi.org/10.1364/OL.43.005853

[39] B.M. Anderson, J. Soloman, A. Flores, 1.1 kW, beam-combinable thulium doped all-fibre amplifier, Conf. Fiber Lasers XVIII - Technology and Systems, Proc. SPIE 11665 (2021) 116650B. https://doi.org/10.1117/12.2576209

[40] C. Ren, Y. Shen, Y. Zheng, Y. Mao, F. Wang, D. Shen, H. Zhu, Widely-tuneable all-fibre Tm doped MOPA with > 1 kW of output power, Opt. Ex. 2931 (2023) 22733–22739. https://doi.org/13.1364/OE.494015

[41] M. Michalska, P. Honzatko, P. Grzes, M. Kamradek, O. Podrazky, I. Kasik, J. Swiderski, Thulium-doped 1940-and 2034-nm fibre amplifiers: Towards highly efficient, high-power all-fibre laser systems, J. Lightwave Technol. 42 (2024) 339-346. https://doi.org/10.1109/JLT.2023.3301397

[42] D. Panitzek, L. Romano, M. Eichhorn, C. Kieleck, Optimization of the slope efficiency of a core-pumped thulium-doped fibre laser by the thermally diffused expanded-core technique, Optics Continuum 3 (2024) 732-737. https://doi.org/10.1364/OPTCON.519347

[43] B. Jirickova, M. Grabner, C. Jauregui, J. Aubrecht, O. Schreiber, P. Peterka, Temperature-dependent cross section spectra for thulium-doped fibre lasers, Opt. Lett. 48 (2023) 811-814. https://doi.org/10.1364/OL.479313

[40] D.C. Hanna, R.M. Percival, R.G. Smart, J.E. Townsend, A.C. Tropper, Continuous-wave oscillation of holmium-doped silica fibre laser, Electron. Lett. 25 (1989) 593-594. https://doi.org/10.1049/el:19890403

[45] S.D. Jackson, A. Sabella, A. Hemming, S. Bennetts, D.G. Lancaster, High-power 83 W holmium-doped silica fibre laser operating with high beam quality, Opt. Lett. 32 (2007) 241-243. https://doi.org/10.1364/OL.32.000241

[46] A.S. Kurkov, E.M. Sholokhov, V.B. Tsvetkov, A.V. Marakulin, L.A. Minashina, O.I. Medvedkov, A.F. Kosolapov, Holmium fibre laser with record quantum efficiency, Quantum el. 41 (2011) 492-494. https://doi.org/10.1070/QE2011v041n06ABEH014565

[47] A. Hemming, N. Simakov, j. Haub, A. Carter, A review of recent progress in holmium-doped silica fibre sources, Opt. fibre technol. 20 (2014) 621-630. https://doi.org/10.1016/j.yofte.2014.08.010

[48] N. Simakov, Z. Li, Y. Jung, J.M.O. Daniel, P. Barua, P.C. Shardlow, S. Liang, J.K. Sahu, A. Hemming, W.A. Clarkson, S.-U. Alam, D.J. Richardson, High gain holmium-doped fibre amplifiers, Opt. Ex. 24 (2016) 13946-13956. https://doi.org/10.1364/OE.24.013946

[49] A.V. Kir'yanov, Y.O. Barmenkov, I.V. Garcia, 2017. 2.05 μm holmium-doped all fibre laser diode-pumped at 1.125 μm, Laser Phys. 27, 085101. https://doi.org/10.1088/1555-6611/aa7378

[50] L.G. Holmen, P.C. Shardlow, P. Barua, J.K. Sahu, N. Simakov, A. Hemming, W.A. Clarkson, Tuneable holmium-doped fibre laser with multiwatt operation from 2025 nm to 2200 nm, Opt. Lett. 44 (2019), 4131-4134. https://doi.org/10.1364/OL.44.004131

[51] A. Hemming, N. Simakov, A. Davidson, M. Oermann, L. Corena, D. Stepanov, N. Carmody, J. Haub, R. Swain, A. Carter, Development of high power holmium-doped fibre amplifiers, Conf. Fiber Lasers XI - Technology, Systems, and Applications, Proc. SPIE 8961 (2014) 89611A. https://doi.org/10.1117/12.2042963

[52] B. Beaumont, P. Bourdon, A. Barnini, L. Kervella, T. Robin, J.L. Gouët, High efficiency holmium-doped triple-clad fiber laser at 2120 nm, J. Lightwave Technol. 40 (2022) 6480–6485. https://doi.org/10.1109/JLT.2022.3196807

[53] J.Pokorny, B. Svejkarova, J. Aubrecht, M. Kamradek, I. Barton, I. Kasik, P. Honzatko, P. Peterka, Holmium-doped silica fibre combining high doping and high efficiency, Proc. Optica Laser Congress and Exhibitions (ASSL), Osaka, Oct. 2024.

[54] M.E. Fermann, D.C. Hanna, D.P. Shepherd, P.J. Suni, J.E. Townsend, Efficient operation of an Yb-sensitized Er fibre laser at 1.56 μm, Electron. Lett. 24 (1988) 1135-1136. https://doi.org/10.1049/el:19880772

[55] D.C. Hanna, R.M. Percival, I.R. perry, R.G. Smart, A.C. Tropper, Efficient operation of an Yb-sensitized Er fibre laser pumped in 0.8 μm region, Electron. Lett. 24 (1988) 1068-1069. http://dx.doi.org/10.1049/el:19880724

[56] G.T. Maker, A.I. Ferguson, 1.56 μm Yb-sensitized Er fibre laser pumped by diode-pumped Nd:YAG and Nd:YLF lasers, Electron. Lett. 24 (1988) 1160-1161. https://doi.org/10.1049/el:19880788

[57] G.G. Vienne, J.E. Caplen, L. Dong, J.D. Minelly, J. Nilsson, D.N. Payne, Fabrication and characterization of Yb 3+ : Er 3+ phosphosilicate fibres for lasers, J. Lightwave Technol. 16 (1998) 1990-2001. http://dx.doi.org/10.1109/50.730360

[58] D.C. Hanna, A. Kazer, M.W. Phillips, P.J. Suni, Active mode-locking of an Yb:Er fibre laser, Electron. Lett. 25 (1989) 95-96. https://doi.org/10.1049/el:19890070

[59] A.B. Grudinin, D.J. Richardson, A.K. Senatorov, D.N. Payne, Nd:YAG laser pumped picosecond $Yb^{3+}/Er^{3+}$ fibre laser, Electron. Lett. 28 (1992) 766-767. http://dx.doi.org/10.1049/el:19920484

[60] D.U. Noske, A. Boskovic, M.J. Guy, J.R. Taylor, Synchronously pumped, picosecond, ytterbium-erbium fibre laser, Electron. Lett. 29 (1993) 1863-1864. https://doi.org/10.1049/el:19931240

[61] S.G. Grubb, W.H. Humer, R.S. Cannon, S.W. Vendetta, K.L. Sweeney, P.A. Leilabady, M.R. Keur, J.G. Kwasegroch, T.C. Munks, D.W. Anthon, +24 6 dBm output power Er/Yb codoped optical amplifier pumped by diode-pumped Nd:YLF laser, Electron. Lett. 28 (1992) 1275-1276. https://doi.org/10.1049/el:19920807

[62] Z.J. Chen, J.D. Minelly, Y. Gu, Compact low cost $Er^{3+}$ / $Yb^{3+}$ codoped fibre amplifiers pumped by 827 nm laser diode, Electron. Lett. 32 (1996) 1812-1813. http://dx.doi.org/10.1049/el:19961189

[63] J.T. Kringlebotn, J.L. Archambault, L. Reekie, J.E. Townsend, G.G. Vienne, D.N. Payne, Highly-efficient, low-noise grating-feedback $Er^{3+}$ : $Yb^{3+}$ codoped fibre laser, Electron. Lett. 30 (1994) 972 -973. https://doi.org/10.1049/el:19940628

[64] S. Gray, J.D. Minelly, A.B. Grudinin, J.E. Caplen, 1Watt Er/Yb singlemode superfluorescent optical fibre source, Electron. Lett. 33 (1997) 1382-1383. https://doi.org/10.1049/el:19970925

[65] W. Li, Q. Qiu, L. Yu, Z. Gu, L. He, S. Liu, X. Yin, X. Zhao, J. Peng, H. Li, Er/Yb co-doped 345-W all-fibre laser at 1535 nm using hybrid fibre, Opt. Ex. 48 (2023) 3027–3030. https://doi.org/10.1364/OL.491863

[66] K. Krzempek, G. Sobon, K. M. Abramski, DFG-based mid-IR generation using a compact dual-wavelength all-fibre amplifier for laser spectroscopy applications, Opt. Ex. 21 (2013) 20023–20031. https://doi.org/10.1364/OE.21.020023

[67] R. Buczynski, M. Franczyk, D. Pysz, J. Aubrecht, G. Stepniewski, A. Filipkowski, M. Kamradek, I. Kasik, P. Peterka, Development of active fibres with nanostructured cores, 2023 Conference on Lasers and Electro-Optics Europe & European Quantum Electronics Conference (CLEO/Europe-EQEC), Munich, Germany, 2023, pp. 01-01. https://doi.org/10.1109/CLEO/EUROPE-EQEC57999.2023.10232736

[68] I. Barton, M. Franczyk, P. Peterka, J. Aubrecht, P. Varak, M. Kamradek, O. Podrazky, R. Kasztelanic, R. Buczynski, I. Kasik, 2023. Optimization of erbium and ytterbium concentration in nanostructured core fibre for dual-wavelength fibre lasers, Proc. Specialty Optical Fibres, SPIE 12573, 1257311. https://www.doi.org/10.1117/12.2666703

[69] P. Peterka, I. Barton, P. Varak, M. Franczyk, J. Aubrecht, R. Kasztelanic, O. Podrazky, I. Kasik, R. Buczynski, 2024. Dual-wavelength fibre lasers based on active fibres with structured cores, Proc. Fiber Lasers XXI: Technology and Systems, SPIE 12865, 1286512. https://doi.org10.1117/12.3000800

[70] P. Miluski, K. Markowski, M. Kochanowicz, M. Lodzinski, W.A. Pisarski, J. Pisarska, M. Kuwik, M. Lesniak, D. Dorosz, J. Zmojda, T. Ragin, J. Dorosz, 2023. Broadband profiled eye-safe emission of LMA silica fibre doped with $Tm^{3+}$/$Ho^{3+}$ ions, Materials 16, 7679. https://doi.org/10.3390/ma16247679

[71] P. Honzatko, Y. Baravets, I. Kasik, O. Podrazky, Wideband thulium-holmium-doped fibre source with combined forward and backward ASE at 1600-2300 nm spectral band, Opt. Lett. 39 (2014) 3650-3653. https://doi.org/10.1364/OL.39.003650

[72] N.J. Ramirez-Martinez, M. Nunez-Velazquez, J.K. Sahu, Study on the dopant concentration ratio in thulium-holmium doped silica fibres for lasing at 2.1µm, Opt. Ex. 28 (2020) 24961-24967. https://doi.org/10.1364/OE.397855

[73] R.P. Tuminelli, B.C. McCollum, E. Snitzer, Fabrication of high-concentration RE-doped optical fibres using chelates, J. Lightwave Technol. 8 (1990) 1680-1683. https://doi.org/10.1109/50.60565

[74] A. Martinez, L.A. Zenteno, J.C.K. Kuo, Optical and spectroscopic characterization of Nd-doped aluminosilicate fibre preforms made by the MCVD method using chelate delivery, Appl. Phys., B67 (1998) 17-21. https://doi.org/10.1007/s003400050468

[75] N. Choudhury, N.K. Shekhar, A. Dhar, R. Sen, Graded-index ytterbium-doped optical fibre fabricated through vapor phase chelate delivery technique, Phys. Status Solidi A 216 (2019) 1900365. https://doi.org/10.1002/pssa.201900365

[76] B. Lenardic, M. Kveder, Advanced vapor-phase doping method using chelate precursor for fabrication of rare earth-doped fibres, Proc. Conf. Optical fibre communication (OFC 2009), 1538-1540. https://doi.org/10.1364/OFC.2009.OThK6

[77] A.L.G. Carter, S.B. Poole, M.G. Sceats, Flash-condensation technique for the fabrication of high-phosphorus-content RE-doped fibres, Electron. Lett. 21 (1992) 2009-2010. https://doi.org/10.1364/OAA.1992.PD6

[78] K.S. Park, B.W. Lee, M. Choi, An analysis of aerosol dynamics in the modified chemical vapor deposition, Aerosol Sci. Technol. 31 (1999) 258-274. https://doi.org/10.1080/027868299304147

[79] https://focenter.com/fiber-process-development-manufacturing-technical-papers, 2024 (accessed 4 September 2024)

[80] J. Ballato, A.C. Peacock, 2018. Perspective: Molten core optical fibre fabrication – A route to new materials and applications, APL Photonics 3, 120903. https://doi.org/10.1063/1.5067337

[81] Highly efficient Yb-doped silica fibres prepared by powder sinter technology, Opt. Lett. 36 (2011) 1557-1559. https://doi.org/10.1364/OL.36.001557

[82] K. Schuster, S. Unger, C. Aichtele, F. Lindner, S. Grimm, D. Litzkendorf, J. Kobelke, J. Bierlich, K. Wondraczek, H. Bartelt, Material and technology trends in fibre optics, Adv. Opt. Technol. 3 (2014) 447–468. http://dx.doi.org/10.1515/aot-2014-0010

[83] R. Renner-Erny, L. Di Labio, W. Luthy, A novel technique for active fibre production, Opt. Materials 29 (2007) 919-922. https://doi.org/10.1016/j.optmat.2006.02.004

[84] O. Sysala, I. Kasik, I. Spejtkova, Preparation of preforms and optical fibres containing aluminum by the solution-doping method, Ceramics-Silikaty 35 (1991) 363-367.

[85] V. Matejec, I. Kasik, D. Berkova, M. Hayer, J. Kanka, Sol-gel fabrication and properties of silica cores of optical fibres doped with $Yb^{3+}$, $Er^{3+}$, $Al_2O_3$ or $TiO_2$, J. Sol-gel Sci. and Technol. 13 (1998) 617-621. https://doi.org/10.1023/A:1008655630006

[86] O.V. MAzurin, M.V. Strelcina, T.P. Schvajko-Schvajkovskaya, Properties of glasses, Nauka, Leningrad, 1980.

[87] Y. Ohmori, F. Hanawa, M. Nakahara, Fabrication of low-loss $Al_2O_3$-doped silica fibres, Electron. Lett. 18 (1982) 761-763. http://dx.doi.org/10.1049%2Fel%3A19820515

[88] Y. Ohmori, T. Miya, M. Horiguchi, Transmission-loss characteristics of $Al_2O_3$-doped silica fibre, J. Ligthwave Technol. 1 (1983) 50-56. https://doi.org/10.1109/JLT.1983.1072067

[89] J.R. Simpson, J.B. MacChesney, Optical fibres with an $Al_2O_3$-doped silicate core composition, Electron. Lett. 19 (1983) 261-262. https://doi.org/10.1049/el:19830180

[90] V. Matejec, I. Kasik, M. Pospisilova, Preparation and optical properties of silica optical fibres with an $Al_2O_3$-doped core, J. Non-Cryst Solids 192&193 (1995) 195-198. https://doi.org/WOS:A1995TJ24700043

[91] F. Lindner, A. Kriltz, A. Scheffel, A. Dellith, J. Dellith, K. Wondraczek, H. Bartelt, Influence of process parameters on the incorporation of phosphorus into silica soot material during MCVD process, Opt. Mat. Ex. 10 (2020) 763-773. https://doi.org/10.1364/OME.385056

[92] I. Kasik, V. Matejec, M. Pospisilova, J. Kanka, J. Hora, Silica optical fibres doped with $Yb^{3+}$ and $Er^{3+}$, Proc. Alt '95 International Symposium On Advanced materials for optics and optoelectronics, SPIE 2777 (1996) 71-79. https://doi.org/10.1117/12.232228

[93] B.J. Ainslie, S.P. Craigh, S.T. Davey, B. Wakefield, The fabrication, assessment and optical properties of high-concentration $Nd^{3+}$-and $Er^{3+}$-doped silica-based fibres, Mat. Lett. 6 (1988) 139-144. https://doi.org/10.1016/0167-577X(88)90086-9

[94] B.J. Ainslie, S.P. Craig, S.T. Davey, The absorption and fluorescence spectra of rare earth ions in silica-based monomode fibre, J. Lightwave Technol. 6 (1988) 287-293. https://doi.org/10.1109/50.4001

[95] K. Arai, H. Namikawa, K. Kumata, T. Honda, Aluminum or phosphorus co-doping effects on the fluorescence and structural properties on neodymium-doped silica glass, J. Appl. Phys. 59 (1986) 3430-3436. https://doi.org/10.1063/1.336810

[96] B.J. Ainslie, S.P. Craig, S.T. Davey, The fabrication and optical properties of $Nd^{3+}$ in silica-based optical fibres, Mat. Lett. 5 (1987) 143-146. https://doi.org/10.1016/0167-577X(87)90023-1

[97] M.E. Likhachev, M.M. Bubnov, K.V. Zotov, D.S. Lipatov, M.V. Yaskov A.N. Guryanov, Effect of the $AlPO_4$ join on the pump-to-signal conversion efficiency in heavily Er-doped fibres, Opt. Lett. 34 (2009) 3355-3357. https://doi.org/10.1364/OL.34.003355

[98] M.M. Bubnov, A.N. Guryanov, K.V. Zotov, L.D. Iskhakova, S.V. Lavrishchev, D.S. Lipatov, M.E. Likhachev, A.A. Rybaltovksy, V.F. Khopin, M.V. Yashkov, E.M. Dianov, Optical properties of fibres with aluminophosphosilicate glass cores, Quantum el. 39 (2009) 857-862. https://doi.org/10.1070/QE2009v039n09ABEH014007

[99] M.E. Likhachev, M.M. Bubnov, K.V. Zotov, O.I. Medvedkov, D.S. Lipatov, M.V. Yashkov, A.N. Guryanov, Erbium-doped aluminophosphosilicate optical fibres, Quantum. el. 40 (2010) 633-638. https://doi.org/10.1070/QE2010v040n07ABEH014326

[100] S. Jetschke, S. Unger, A. Schwuchow, M. Leich, J. Kirchhof, Efficient Yb laser fibres with low photodarkening by optimization of the core composition, Opt. Ex. 16 (2008) 15540-15545. https://doi.org/10.1364/oe.16.015540

[101] M. Ferrari, G.C. Righini, Glass-ceramic materials for guided-wave optics, Int. J. Appl. Glass Sci. 6 (2015) 240-248. https://doi.org/10.1111/ijag.12129

[102] A. Lin, S. Boo, D.S. Moon, H. Jeong, Y. Chung, W.T. Han, Luminescence enhancement by Au nanoparticles in $Er^{3+}$-doped germano-silicate optical fibre, Opt. Ex. 15 (2007) 8603-8608. https://doi.org/10.1364/oe.15.008603

[103] P.R.Watekar, S. Ju, W.T. Han, Optical properties of the alumino-silicate glass doped with Er-ions/Au particles, Colloids and surfaces A: Physicochem. and Eng. Aspects 313 (2008) 492–496. https://doi.org/10.1016/J.COLSURFA.2007.04.178

[104] W. Blanc, Z. Lu, T. Robine, F. Pigeonneau, C. Molardi, D. Tosi, Nanoparticle in optical fibre, issue and opportunity of light scattering (invited), Opt. Mat. Ex. 12 (2022) 2635-2652. https://doi.org/10.1364/OME.462822

[105] C.C. Baker, E.J. Friebele, A.A. Burdett, D.L. Rhonehouse, J. Fontana, W. Kim, S.R. Bowman, L.B. Shaw, J. Sanghera, J. Zhang, R. Pattnaik, M. Dubinskii, J. Ballato, C. Kucera, A. Vargas, A. Hemming, N. Simakov J. Haub, Nanoparticle doping for high power fibre lasers at eye-safer wavelengths, Opt. Ex. 25 (2017) 13903-13915. https://doi.org/10.1364/OE.25.013903

[106] J. Koponen, L. Petit, K. Teemu, 2011. Progress in direct nanoparticle deposition for the development of the next generation fibre lasers, Opt. Eng. 50, 111605. http://dx.doi.org/10.1117/1.3613944

[107] M.C. Paul, S. Bysakh, S. Das, M. Pal, S.K. Bhadra, S. Yoo, A.J. Boyland, J.K. Sahu, Nano-engineered $Yb_2O_3$ doped optical fibre: fabrication, material characterizations, spectroscopic properties and lasing characteristics: a review, Sci. Adv. Materials 4 (2012) 292-321. https://doi.org/10.1166/sam.2012.1284

[108] A. Dhar, I. Kasik, B. Dussardier, O. Podrazky, V. Matejec, Preparation and properties of Er–doped $ZrO_2$ nanocrystalline phase-separated preforms of optical fibres by MCVD process, Int. J. Appl. Cer. Technol. 9 (2012) 341-348. https://doi.org/10.1111/j.1744-7402.2011.02669.x

[109] D. Ghosh, N. Choudhury, S. Balaji, K. Dana, A. Dhar, Synthesis and characterization of $Tm_2O_3$-doped $Lu_2O_3$ nanoparticle suitable for fabrication of thulium-doped laser fibre, J. Mat. Sci.: Materials in Electronics 32 (2021) 4505-4514. https://link.springer.com/article/10.1007/s10854-020-05191-9

[110] A. Veber, Z. Lu, M. Vermillac, F. Pigeonneau, W. Blanc, Nanostructured optical fibres mase of glass-ceramics, and phase separated and metallic particle-containing glasses, Fibres 7 105 (2019) 1-29. https://doi.org/10.3390/fib7120105

[111] B. Lenardic, M. Kveder, D Lisjak, H. Guillion, S. Bonnafous, Novel method for fabrication of metal- or oxide-nanoparticle doped silica-based specialty optical fibres, Proc. Conf. Optical Components and Materials VIII, SPIE 7934, 2011. http://dx.doi.org/10.1117/12.871873

[112] I. Kasik, P. Peterka, J. Mrazek, P. Honzatko, Silica optical fibres doped with nanoparticles for fibre lasers and broadband sources, Current Nanoscience 12 (2016) 277-290. http://dx.doi.org/10.2174/1573413711666150624170638

[113] J. Ballato, H. Ebendorf-Heidepriem, J. Zhao, L. Petit, Glass and process development for the next generation of optical fibres: a review, Fibres 5, (2017) 1-25. https://doi.org/10.3390/fib5010011

[114] P.D. Dragic, M. Cavillon, J. Ballato, 2018. Materials for optical fibre lasers: a review, Appl. Phys. Reviews 5, 041301. https://doi.org/10.1063/1.5048410

[115] O. Podrazky, I. Kasik, M. Pospisilova, V. Matejec, Use of alumina nanoparticles for preparation of erbium-doped fibres, IEEE LEOS Annual Meeting Conference Proc. 1-2 (2007) 246-247. https://doi.org/10.1109/LEOS.2007.4382369

[116] O. Podrazky, I. Kasik, M. Pospisilova, V. Matejec, Use of nanoparticles for preparation of rare-earth doped silica fibres, Phys. Status Solidi C - Current Topics In Solid State Phys. 6 (2009) 2228-2230. https://doi.org/10.1002/pssc.200881727

[117] M. Kamradek, I. Kasik, J. Aubrecht, J. Mrazek, O. Podrazky, J. Cajzl, P. Varak, V. Kubecek, Ceramic nanoparticle-doping implementation into MCVD method for fabrication of holmium-doped fibres for fibre lasers, IEEE Phot. J. 11 (2019) 1-10, https://doi.org/10.1109/JPHOT.2019.2940747

[118] C.C. Baker, A. Burdett, E.J. Friebele, D.L. Rhonehouse, W. Kim, J. Sanghera, Rare earth co-doping for increased efficiency of resonantly pumped Er-fibre lasers, Opt. Mat. Ex. 9 (2019) 1041-1048. https://doi.org/10.1364/OME.9.001041

[119] E. Mobini, S. Rostami, M. Peysokhan, A. Albrecht, S. Kuhn, S. Hein, C. Hupel, J. Nold, N. Haarlammert, T. Schreiber, R. Eberhardt, A. Tunnermann, M. Sheik-Bahae, A. Mafi, Commun. Phys. 3 (2020) 134. https://doi.org/10.1038/s42005-020-00401-6

[120] W.J. Wadsworth, R.M. Percival, G. Bouwmans, J.C. Knight, P.S. J. Russell, High power air-clad photonic crystal fibre laser, Opt. Ex. 11, (2003) 48–53. https://doi.org/10.1364/OE.11.000048

[121] J. Hecht, City of light: the story of fibre optics, Oxford University Press, 1999, New York, ISBN 0-19-510818-3; 0-19-516255-2.

[122] P. Peterka, J. Aubrecht, D. Pysz, M. Franczyk, O. Schreiber, M. Kamradek, I. Kasik, R. Buczynski, Development of pedestal-free large mode area fibres with $Tm^{3+}$ doped silica nanostructured core, Opt. Ex. 31 (2023) 43004-43016. https://doi.org/10.1364/OE.503047

[123] M. Franczyk, D. Pysz, R. Stepien, J. Cimek, R. Kasztelanic, F. L. Chen, M. Klimczak, L. Zhao, I. Kasik, P. Peterka, R Buczynski, Dual band active nanostructured core fibre for two-color fibre laser operation, J. Lightwave Technol. 40 (2022) 7180 – 7190. https://doi.org/10.1109/JLT.2022.3199581

[124] P. Varak, I. Kasik, P. Peterka, J. Aubrecht, J. Mrazek, J. Kamradek, O. Podrazky, I Barton, M. Franczyk, R. Buczynski, P. Honzatko, Heat treatment and fibre drawing effect on the luminescence properties of RE-doped optical fibres (RE = Yb, Tm, Ho), Opt. Ex. 30 (2022) 10050-10062. https://doi.org/10.1364/OE.449643

[125] P. Varak, M. Kamradek, J. Mrazek, O. Podrazky, J. Aubrecht, P. Peterka, I. Kasik, 2022. Luminescence and laser properties of RE-doped silica optical fibres: the role of composition, fabrication processing, and inter-ionic energy transfers, Opt. Materials: X 15, 100177. https://doi.org/10.1016/j.omx.2022.100177

[126] P. Varak, M. Kamradek, J. Aubrecht, O. Podrazky, J. Mrazek, I. Barton, A. Michalcova, M. Franczyk, R. Buczynski, I. Kasik, P. Peterka, P. Honzatko, Heat treatment and fibre drawing effect on the matrix structure and fluorescence lifetime of Er and Tm-doped silica optical fibres, Opt. Materials Ex. 14 (2024) 1048 – 1061. https://www.doi.org/10.1364/OME.520422

[127] J.E. Townsend, W.I. Barnes, K.P. Jedrzejewski, S.G. Grubb, $Yb^{3+}$ sensitized $Er^{3+}$ doped silica optical fibre with ultrahigh transfer efficiency and gain, Electron. Lett. 27 (1991) 1958-1959. https://doi.org/10.1049/EL%3A19911214

[128] I. Kasik, Data for Doped and structured silica optical fibres for fibre laser sources (2024). https://doi.org/10.5281/zenodo.13831989